\documentclass{article}
\usepackage[left=1.1in,right=1.1in,top=.95in,bottom=.95in]{geometry}
\usepackage[utf8]{inputenc} %
\usepackage[T1]{fontenc}    %
\usepackage{amsmath}
\usepackage{amssymb}
\usepackage{amsthm}
\usepackage{mathtools}
\usepackage{bbm}
\usepackage[mathscr]{euscript}
\usepackage{graphicx}
\usepackage{subcaption}
\usepackage{float}
\usepackage{tikz}
\usetikzlibrary{decorations.pathreplacing}
\usepackage{xcolor}
\usepackage{hwemoji}
\usepackage{footnote}  %
\usepackage{algorithm2e}
\usepackage{multicol}
\usepackage{dialogue}
\usepackage{csquotes}
\usepackage{cancel}
\usepackage{placeins}  %
\usepackage[bottom]{footmisc}
\usepackage[affil-it]{authblk}
\definecolor{linkcolor}{HTML}{0050FF}
\definecolor{citecolor}{HTML}{0050FF}
\definecolor{urlcolor}{HTML}{0050FF}
\usepackage[breaklinks,colorlinks=true,
            citecolor=citecolor,urlcolor=urlcolor,linkcolor=linkcolor]{hyperref}
\usepackage[noabbrev,capitalize]{cleveref}
\usepackage[
    backend=biber,
    citestyle=ext-authoryear-comp,
    uniquename=false,
    bibencoding=utf8,
    sorting=nty,
    sortcites=false
    doi=false,isbn=false,url=false,
    maxcitenames=2,
    uniquelist=false,
    maxbibnames=123,
    backref=true,
    hyperref=true,
]{biblatex}
\usepackage{setspace}
\usepackage{microtype}
\usepackage{tcolorbox}
\tcbuselibrary{theorems,skins}

\DeclareFieldFormat[article,inproceedings,manual,book,inbook,misc,thesis,incollection]{title}{%
  \iffieldundef{url}{#1}{\href{\thefield{url}}{#1}}}

\DeclareInnerCiteDelims{cite}{(}{)}

\newcommand{\blue}[1]{\textcolor{blue}{#1}}
\newcommand{\red}[1]{\textcolor{red}{#1}}

\newcommand{\Z}{\mathbb{Z}}
\newcommand{\G}{\mathcal{G}}
\newcommand{\V}{\mathcal{V}}
\newcommand{\E}{\mathcal{E}}
\newcommand{\T}{\mathbb{T}}
\newcommand{\X}{\mathbb{X}}
\newcommand{\Y}{\mathbb{Y}}

\newcommand{\Tcal}{\mathcal{T}}
\renewcommand{\S}{\mathcal{S}}
\renewcommand{\P}{\mathcal{P}}
\newcommand{\Xt}{X_{\T_x}}
\newcommand{\Yt}{Y_{\T_y}}
\newcommand{\Zt}{Z_{\T_z}}

\newcommand{\tick}{\boldsymbol{\circ}}
\newcommand{\tickx}{\boldsymbol{\times}}
\renewcommand{\t}{\bar{t}}

\newif\ifanonymous
\anonymousfalse

\newtheorem{remark}{Remark}

\newcounter{example}
\newenvironment{example}[2][]{%
  \refstepcounter{example} %
  \paragraph{Example~\theexample#2.}%
}{}
\newenvironment{example*}[2][]{%
  \paragraph{Example #2.}
}{}

\newtcolorbox[]{mydefinition*}[1][]{%
    colback=blue!5,    %
    colframe=blue!20,   %
    coltitle=black,     %
    fonttitle=\bfseries, %
    sharp corners,      %
    title=#1
}

\newtcbtheorem[number within=section]{graydefinition}{Definition}{
    colback=black!5,
    colframe=black!20, 
    coltitle=black, 
    sharp corners, 
    fonttitle=\bfseries,
    label type=graydefinition}{graydef}
\crefname{graydefinition}{Definition}{Definitions}

\title{The Case for Time in Causal DAGs}
\author{}
\date{}

\makeatletter
\renewenvironment{abstract}{
    \par\noindent\begin{center}\textbf{\abstractname}\ \end{center}}
    {\par\bigskip}
\makeatother

\begin{document}

\maketitle

\vspace{-3.5em}

\ifanonymous

\else
\begin{center}
\begin{minipage}{0.9\linewidth}
\begin{center}
\begin{multicols}{2}
    \textbf{Alexander G. Reisach}\footnotemark\\
    CNRS, MAP5\\
    Université Paris Cité\\
    F-75006 Paris, France
    
    \columnbreak

    \textbf{Alberto Suárez}\\   
    Departamento de Ingeniería Informática,\\
    Escuela Politécnica Superior,\\    
    Universidad Autónoma de Madrid, Spain
\end{multicols}
\begin{multicols}{2}

    \textbf{Sebastian Weichwald}\\
    Department of Mathematical Sciences and\\
    Pioneer Centre for Artificial Intelligence,\\
    University of Copenhagen, Denmark
    
    \columnbreak

    \textbf{Antoine Chambaz}\\   
    CNRS, MAP5\\    
    Université Paris Cité\\ 
    F-75006 Paris, France   
\end{multicols}
\end{center}
\end{minipage}
\end{center}
\footnotetext{Correspondence to \url{alexander.reisach@math.cnrs.fr}}
\fi

\vskip1.5em
\begin{abstract}
\begin{center}
\noindent\begin{minipage}{0.75\linewidth}
We make the case for incorporating a notion of time into causal directed acyclic graphs (DAGs).
We demonstrate that nontemporal causal DAGs are ambiguous and obstruct justification of the acyclicity assumption.
Assuming that causes precede effects, causal relationships are relative to the time order, and causal DAGs require temporal qualification.
We propose a formalization via \textit{composite} causal variables that refer to quantities at one or multiple time points.
We emphasize that the acyclicity assumption requires different justifications depending on whether the time order allows cycles.
We conclude by discussing implications for the interpretation and applicability of DAGs as causal models.
\end{minipage}
\end{center}
\end{abstract}
\vskip-1.em
\hfill
\section{Nontemporal Causal DAGs}
Causal directed acyclic graphs (DAGs), as popularized by \cite{pearl2000causality,pearl2009causality}, 
have been conceived as nontemporal causal models.
\textcite[Section 7.5.1]{pearl2009causality} states on the matter of using time to determine causation:
\begin{quote}
    \vskip-.5em
    ``The reliance on temporal information has its price, as it excludes a priori the analysis of cases in which the temporal order is not well-defined.''
\end{quote}
\vskip-.5em
Time is real, however, and there is no escaping it and whatever price it really may exact, whether temporal information is relied upon or not.
We therefore argue that it is best to use time explicitly, and set out to explore the limitations and opportunities it reveals.
In doing so, we rely on the principle that
\textit{causes precede effects}.
This principle is fundamental to the definition of causality.
In proposing the method of path coefficients, the precursor to graphical causal models, \cite{wright1934method} states that 
``only such paths are tried which are appropriate in direction in time''.
The Bradford Hill criteria for causality \parencite{hill1965environment}, widely used in epidemiology, refer to the same principle as ``temporality''.
\cite{Granger1969}, \cite{Suppes1970}, and \cite{shoham1990nonmonotonic} rely on the time order between cause and effect for their theories of causality.
Throughout their foundational work on causal discovery, \cite{spirtes1993causation,spirtes2001causation} emphasize the existence of a time order.
Similarly, the canonical work on graphical causal models \textcite[][e.g.\ Section 7.5.1]{pearl2009causality}, as well as \textcite[][e.g.\ Chapter 10]{peters2017elements}, concur that causality follows the time order.

Despite its central role in causality, the notion of time is absent from the formal treatment of causal DAGs, which define variables and edges without an explicit reference to time.
In contrast to the potential outcomes framework of statistical causality 
\parencite{neyman1932,rubin1974estimating},
causal DAGs have been developed explicitly for ``nontemporal statistical data'' \parencite[][Chapter~2]{pearl2009causality}.
This expression covers two distinct properties.
First, the distributions compatible with a DAG do not depend on time.
Second, and perhaps more controversially, a DAG imposes no constraints on the temporal structure within realizations of such a distribution beyond those entailed by the causal order
(see \cref{sec:formal_time} for details).
This offers a powerful unified theory regardless of whether the time order prevents cycles, or whether they are possible but absent \parencite[][Section 7.5.1]{pearl2009causality}.
However, the general nature of this approach obscures the connection between time and causation, as well as the rationale for the acyclicity assumption.
As a result, causal DAGs have been used widely, including in contexts such as neuroimaging \parencite{ramsey2017million} or gene regulatory networks \parencite{lee2019scaling}, where acyclicity is questionable.
In many cases, the search for nontemporal acyclic causal relationships has led to variables without physically well-defined interventions, such as raising cities or changing their climate in the popular \mbox{\textit{altitude} $\to$ \textit{temperature}} example \parencite[e.g.][Section 2.1]{peters2017elements}.
The absence of time has also prompted the use of time proxy variables like \textit{age} or \textit{hour of day} \parencite[e.g.][]{mooij2016distinguishing}, and even the modelling of time itself as a cause \parencite{huang2015identification}, despite the obvious issue of defining corresponding interventions.

These problems are symptomatic of the tension that arises from the contrast between the essential role of time in causation and the nontemporality of causal DAGs:
it is unclear whether a given nontemporal DAG relates quantities at all, any, or some specific time points, and hence how acyclicity should be justified.
Moreover, many real-world variables of interest such as \textit{economic inflation}, \textit{precipitation}, or \textit{clinical depression}, are complex temporal aggregations.
For these reasons, we make the case for incorporating the notion of time into the definition of variables in causal DAGs in an explicit and principled fashion.

\subsection{Nontemporal Causal DAGs are Ambiguous}\label{sec:ambiguity}
In causal DAGs, the values taken by the random variables refer to quantities of a type of object in some external system modelled by the DAG. In this sense, we say as a shorthand that a random variable \textit{refers to} or \textit{measures} quantities.
We refer to a full assignment of values to all random variables of the model as a \textit{realization}.
Consider the DAG $A\to H$ between a random variable $A$ measuring the intake of the drug aspirin
and a random variable $H$ measuring the symptom headache, a popular example in the literature \parencite[see e.g.][]{shpitser2008complete,imbens2015causal,pearl2018book}.
The present edge encodes that $(i)$ \textit{$A$ causes $H$}, 
and the absent edge in the opposite direction encodes that $(ii)$ \textit{$H$ does not cause $A$}.
We rely on the interventionist view of causation \parencite[][]{woodward2005making} typically used in connection with causal DAGs 
\parencites[][Section 4]{pearl1995causal}{pearl2009causality}.
Since causes precede effects, the interventionist reading of $(i)$ entails the statement
\begin{itemize}
    \item[] (I)\quad \textit{\textbf{first} intervening on $A$ may change the \textbf{subsequent} distribution of $H$}
\end{itemize} 
for some unspecified time distance.
However, $ii)$ does not necessarily entail that
\begin{itemize}
    \item[] (II)\quad \textit{\textbf{first} intervening on $H$ does not change the \textbf{subsequent} distribution of $A$}.
\end{itemize}
The ambiguity arises because it is unclear whether nontemporal DAGs make any statement about time orders that do not follow the causal order.
To illustrate this point, consider the two different timelines of instantiated (i.e.\ time point-specific) aspirin and headache variables shown in \cref{fig:illu_timeline}.
\begin{figure}[H]
    \centering
    \begin{subfigure}{.49\linewidth}
        \centering
        \begin{tikzpicture}[scale=1]
            \draw[->,gray,thick] (0,0) -- (3.5,0);
            \draw[gray,thick] (0,0.2) -- (0,-0.2);
            \node at ( 0,0-0.4) {$t_0$};
            \node at ( 3,0-0.4) {$t_1$};
            \node[blue] (x1) at (0,0) {$\tick$};
            \node (X1) at (0,0.4) {$A_{t_0}$};
            \node[red] (y1) at (3,0) {$\tick$};
            \node (Y1) at (3,0.4) {$H_{t_1}$};
            \node (time) at (4,0) {time};
        \end{tikzpicture}
        \caption{Aspirin before headache.}
        \label{fig:tl1}
    \end{subfigure}
    \hfill
    \begin{subfigure}{.49\linewidth}
        \centering
        \begin{tikzpicture}[scale=1]
            \draw[->,gray,thick] (0,0) -- (3.5,0);
            \draw[gray,thick] (0,0.2) -- (0,-0.2);
            \node at ( 0,0-0.4) {$t'_0$};
            \node at ( 3,0-0.4) {$t'_1$};
            \node[blue] (x1) at (3,0) {$\tick$};
            \node (X1) at (3,0.4) {$A_{t'_1}$};
            \node[red] (y1) at (0,0) {$\tick$};
            \node (Y1) at (0,0.4) {$H_{t'_0}$};
            \node (time) at (4,0) {time};
        \end{tikzpicture}
        \caption{Headache before aspirin.}
        \label{fig:tl2}
    \end{subfigure}
    \caption{Timeline of instantiated aspirin (\blue{$\tick$}) and headache (\red{$\tick$}) variables.}
    \label{fig:illu_timeline}
\end{figure}
\noindent
It follows from the causal order that the DAG models a time order of aspirin before headache, as shown in \cref{fig:tl1}, but it is unclear whether it also models the inverse order of headache before aspirin, as shown in \cref{fig:tl2}.
If the DAG is taken to describe only aspirin before headache, then the absence of an effect of $H$ on $A$ follows purely from the time order, and the DAG provides no information about statement (II).
If it is taken to describe both time orders, then the absent edge does entail statement (II) for some unspecified time distance.
In light of the ambiguity, it may seem tempting to interpret the causal order in $A\to H$ as an implicit instantiation of the form $(A_{t_0},H_{t_1})$ with $t_0<t_1$. However, such an interpretation would remain ambiguous for variables not connected by a directed causal path.
Thus, since causal effects depend on time order and may also depend on time distance, the interpretation of nontemporal DAGs is inherently ambiguous.

\subsection{Justifying the Acyclicity Assumption Requires a Notion of Time}\label{sec:acyclicity}
In a nontemporal DAG, it is unclear which time point or set of time points is involved in the definition of each of the random variables.
This issue is particularly problematic when it comes to evaluating the validity of the acyclicity assumption. We summarize the relationship between time order and acyclicity in \cref{remark:cycles}, which is a direct consequence of the principle that causes precede effects within realizations.
\begin{remark}{}
    A causal cycle between two variables requires that at least one of them refers to quantities at multiple time points.
    \label{remark:cycles}
\end{remark}
For illustration, consider a version of the running aspirin and headache example with multiple possible measurements for each, as shown in \cref{fig:justifying_acyclicity}. Assume that aspirin causes subsequent headache and vice versa (aspirin relieves headaches, so the presence of a headache would be a reason to take aspirin).
If $A$ and $H$ each refer to a quantity at a single time point, the DAG between them is acyclic by the time order.
A cycle is possible if one or more variables refer to quantities at multiple time points. We call such variables \textit{composite} variables.
Composite variables can refer to quantities at randomly drawn time points, and they can aggregate quantities at multiple, possibly random, time points.
For instance, $A$ may be uniformly drawn from $\{A_{t_0},A_{t_2}\}$, or be the sum $A_{t_0}+A_{t_2}$.
\begin{figure}[H]
    \centering
    \begin{tikzpicture}[scale=1]
        \node (t) at (11.2,0) {time};
        \draw[->,gray,thick] (0,0) -- (10.5,0);
        \draw[gray,thick] (0,0.2) -- (0,-0.2);
        \draw[gray] (0,0+0.1)--(0,0-0.1);
        \node at ( 0,0-0.4) {$t_0$};
        \node at ( 4,0-0.4) {$t_1$};
        \node at ( 6,0-0.4) {$t_2$};
        \node at (10,0-0.4) {$t_3$};
        \node[blue] (x1) at (0,0) {$\tick$};
        \node[blue] (x2) at (6,0) {$\tick$};
        \node[red] (y1) at (4,0) {$\tick$};
        \node[red] (y1) at (10,0) {$\tick$};
        \node (X1) at (0,0.4) {$A_{t_0}$};
        \node (X1) at (6,0.4) {$A_{t_2}$};
        \node (X1) at (4,0.4) {$H_{t_1}$};
        \node (X1) at (10,0.4) {$H_{t_3}$};
    \end{tikzpicture}
    \vskip-.5em
    \caption{A timeline with multiple possible measurement times for aspirin and headache.}
    \label{fig:justifying_acyclicity}
\end{figure}
\noindent
More generally, the example highlights that a DAG cannot relate the generic concepts of aspirin and headache, but only temporal instantiations thereof.
For example, the DAG may change from $A\to H$ for $A\coloneq A_{t_0}$ and $H\coloneq H_{t_1}$, to $H\to A$ for $H\coloneq H_{t_1}$ and $A\coloneq A_{t_2}$, to the cyclic causal graph
\hspace{-1.23em}
\tikz[baseline=(base)]{
\begin{tikzpicture}
    \node (base) at (0,-0.05) {}; %
    \node (uX) at (0,.05) {$\phantom{X}$};
    \node (uY) at (1,.05) {$\phantom{Y}$};
    \node (lX) at (0,-0.05) {$\phantom{X}$};
    \node (lY) at (1,-0.05) {$\phantom{Y}$};
    \node (X) at (0,0) {$A$};
    \node (Y) at (1,0) {$H$};
    \draw[->] (uX)--(uY);
    \draw[->] (lY) -- (lX);
\end{tikzpicture}
}$\phantom{a}$
if the variables are composed of multiple time points.
It is thus unclear what causal statements nontemporal DAGs actually imply, and whether acyclicity follows from the time order or is an additional assumption.
To address this issue, we propose a formalism to specify the temporality of variables in causal DAGs.

\subsection{Contribution}
We discuss related work on time in causal modelling and causal time series models in \cref{sec:rel_lit}.
We propose a new formalization of time for causal variables of varying temporal complexity in \cref{sec:formal_time}, and characterize two separate conditions for acyclicity: acyclicity by a time order that prevents cycles, and acyclicity by the absence of causal cycles where the time order allows them.
In \cref{sec:examples} we illustrate our formalism in different temporal configurations, and demonstrate how it can be used to reason systematically about causal cycles, appropriate time scales, and potential faithfulness violations.
In \cref{sec:discussion} we discuss the implications for causal discovery, effect estimation, and the applicability of causal DAGs. \cref{sec:conclusion} summarizes our conclusions,
and \cref{app:formalization} provides additional detailed formalizations.
\section{Related Work}\label{sec:rel_lit}

\paragraph{Causal DAGs.}
DAGs are the central tool for representing causal relationships using graphical models \parencite{pearl2009causality}.
A DAG $\G=(\V, \E)$ consists of a set of vertices $\V$, and a set of directed edges $\E \subseteq \V \times \V$ such that if $(X, Y) \in \E$, then there is a directed edge from $X$ to $Y$.
We call a \textit{directed path} an ordered collection of directed edges such that the end point of the $k$-th edge is the start point of the $(k+1)$-th edge.
By definition, DAGs are acyclic, that is, there are no directed paths that begin and end in the same vertex.
Each vertex represents a random variable, and their joint distribution is assumed to be Markov with respect to the DAG 
\parencites(e.g.)()[][Definition~1.2.2]{pearl2009causality}[][Definition~6.21]{peters2017elements}.
A causal DAG represents an external causal system \parencite[][Section 5]{dawid2010beware}, and a directed edge $(X, Y) \in \E$, also denoted as $X\to Y$, indicates that $X$ is a direct cause of $Y$ within the variables of the graph.
Hence, a DAG encodes \textit{causal structure}, and a DAG with a set of additional parameters is a \textit{causal model} \parencite[][Definitions 2.2.1 and 2.2.2]{pearl2009causality}. 
The superclass of \textit{structural causal models} \parencite[][Definition 7.1.1]{pearl2009causality} does not in principle require acyclic structures, but is nonetheless most often used in conjunction with DAGs 
\parencites(see e.g.)()[][Section 1.2]{pearl2009causality}[][Section 6.2]{peters2017elements},
which are the central object of study and interest in the field of graphical causal models due to their favorable properties.

\paragraph{Critical perspectives on causal DAGs}
Conventionally, causal DAGs relate random variables, which are \textit{type}
variables, whereas the causal order operates within the process by which
these variables are realized, that is, at the \textit{token} level.
For details on the type vs.\ token debate, see \textcite[][Section 2.3]{williamson2004bayesian}, \textcite[][Chapter 14]{spohn2012laws}, and \textcite{spohn2019reply}.
The common justification of assuming the existence of a time order for the acyclicity of a causal DAG implicitly elevates a token-level property to the type-level.
Our work provides the motivation and formalism for aligning the temporal restrictions that act on the token-level with the definition of type-level variables in a principled fashion.

Our contribution ties into a series of critical perspectives on causal DAGs.
Causal DAGs allow for far-reaching causal insights because they are a restrictive model class.
In his methodological critique, \cite{dawid2010beware} points out that the strong assumptions underlying the model class require adequate justification.
\cite{greenland2010overthrowing} argues that causal DAGs rely on null-hypotheses that are implausible in many applications.
\cite{aalen2016can} question whether DAGs between measurements taken at discrete time points are useful for understanding underlying continuous causal processes.
Notably, the works by \cite{greenland2010overthrowing} and \cite{aalen2016can} do not question the directionality of edges or the acyclicity assumption. Writing from the perspective of the applied sciences, they naturally assume that the causal variables refer to quantities measured at certain time points, giving rise to a time order.
This contrasts with the idea of nontemporal \textit{equilibrium} variables predominant in the causal DAG literature 
\parencites(see e.g.)()[][Chapter 2]{pearl2009causality}[][Section 5]{spirtes2010automated}[][Remark 6.7]{peters2017elements}.
\textcite[Section 3.6]{spohn2010structural} notes the problem that nontemporal variables can refer to items of varying temporal positions.
Motivated by the same observation, \cite{weinberger2019reintroducing} proposes a temporal interpretation for equilibrium causal DAGs, and \cite{weinberger2023intervening} discusses the temporality of interventions.
We adopt the applied sciences' pragmatic view that measurements are necessarily time-specific, and take Spohn's critique to its logical conclusion: the causal relationships encoded in DAGs depend on the temporal nature of the variables.
This motivates our development of a formalism that can encode a wide range of possible temporal configurations, highlights the need to justify acyclicity, and clarifies the interpretation of causal DAGs.

\paragraph{Temporal graphical causal models}

There are, of course, many approaches in the literature on graphical causal models that already incorporate a notion of time.
The principal difference of our contribution is that we do not propose a temporal approach as a special or alternative case, but rather argue that all causal DAGs are temporal in nature if one accepts that causes precede effects.

In the existing literature, cyclic graphical causal models \parencite[][]{bongers2021foundations} offer a way to essentially preserve the nontemporal approach while acknowledging that variables that aggregate quantities at multiple time points may be characterized by cyclic causal relationships.
\cite{stern2021causal} points out that considering the temporal order may greatly simplify causal discovery.
A rich literature \parencite[e.g.][]{peters2013causal,malinsky2018causal,runge2018causal,assaad2021time} applies dynamic Bayesian networks \parencite{dean1989model,murphy2002dynamic} as causal models for the classical time series setting involving regular measurements of a dynamic process.
Similarly, \cite{Hansen2014} show that stochastic differential equations can be seen as a continuous-time generalization of time series DAGs. 
Our approach constitutes a bridge between nontemporal and time series causal DAGs.
We highlight that, even if realizations are mutually independent and no matter the level of aggregation, causal variables always have a time dimension which is crucial for interpreting causal relationships and for assessing the validity of the acyclicity assumption.
Conceptually, the perspective we propose is closest to that of the potential outcomes framework \parencite{neyman1932,rubin1974estimating}, and aligns with the definition of causal variables in \textcite[][Section 1.2]{imbens2015causal} and \textcite[][Section 22.4]{hernan2020causal}. 
Our contribution extends the existing literature by considering variables that can refer to quantities at multiple and varying time points. In doing so, we make use of and provide a temporal perspective on the idea behind \textit{cluster} and \textit{group} variables as studied in \cite{anand2023causal} and \cite{wahl2024foundations}.  

\section{Time in Causal DAGs}\label{sec:formal_time}

We set out to formalize a notion of time that is sufficiently flexible to capture variables of different temporal complexity.
To see that there is no contradiction between temporal variables and time-invariant causal mechanisms, 
consider that a realization of the variables in a causal DAG in the sense of \cite{pearl2009causality} results from applying deterministic causal mechanisms to a draw of exogenous noise variables \parencite[following the conception of][]{laplace1814essai}. Each variable in a DAG is typically equipped with its own independent noise source, so the resulting model is still indeterministic.
The causal mechanisms are invariant to the starting time of their application, yet their causal effects unfold in time within each realization. 
We refer to the timeline relative to the starting point of one realization as \textit{within-realization}, and distinguish it from the \textit{between-realization} timeline relating the starting points of different realizations.
For there to be a causal effect between two variables in distribution, their within-realization time order must follow their causal order with positive probability.

To illustrate the distinction between the within-realization and between-realization timelines, \cref{fig:sample_time} provides an example of four realizations of two variables $X$ and $Y$. Each realization contains one value per variable that refers to a quantity at a time point marked by $\tickx$.
This corresponds to a setting where the variables $X$ and $Y$ can refer to one of multiple instantiated variables (that is, variables that refer to a quantity at a specific time point) with positive probability.
\begin{figure}[H]
    \centering
    \begin{subfigure}{\linewidth}
        \centering
        \begin{tikzpicture}[scale=1]
            \node (within) at (1.75,-3.9) {within-realization time};
            \node[rotate=90] (between) at (-.5,-1.5) {between-realization time};
            \draw (3.85,-1.115) rectangle (4.16,-1.425);
            \node[blue] (measurement1) at (4, -1.275) {$\tickx$};
            \node (measurement1) at (4.5, -1.25) {$X$};
            \draw (3.85,-1.615) rectangle (4.16,-1.925);
            \node[red] (measurement1) at (4, -1.775) {$\tickx$};
            \node (measurement1) at (4.5, -1.75) {$Y$};
            \draw[->,thick,dotted] (0,-3.5)--(0,.6);
            \draw[->,thick,dotted] (0,-3.5)--(3.5,-3.5);
            \draw[->,gray,thick] (0.5, 0)--(3,-0);
            \draw[gray,thick]      (0.5,-0.2)--(.5,0.2);
            \node[blue]   (tick1) at (0.5,0) {$\tickx$};
            \node[red]  (tick2) at (2.5,0) {$\tickx$};
            \draw[->,gray,thick] (0.5, -1)--(3,-1);
            \draw[gray,thick]      (0.5,-0.8)--(.5,-1.2);
            \node[blue]  (tick1) at (0.5,-1.0) {$\tickx$};
            \node[red]   (tick2) at (2,-1) {$\tickx$};
            \draw[->,gray,thick] (0.5, -2)--(3,-2);
            \draw[gray,thick]    (0.5,-1.8)--(0.5,-2.2);
            \node[blue]  (tick2) at (2.5,-2) {$\tickx$};
            \node[red]   (tick1) at (0.5,-2) {$\tickx$};
            \draw[->,gray,thick] (0.5, -3)--(3,-3);
            \draw[gray,thick]    (0.5,-2.8)--(0.5,-3.2);
            \node[blue]   (tick2) at (2,-3) {$\tickx$};
            \node[red]  (tick1) at (0.5,-3) {$\tickx$};
        \end{tikzpicture}
    \end{subfigure}
    \caption{Four realizations with their respective timelines.
    }
    \label{fig:sample_time}
\end{figure}

Note that the causal order restricts the within-realization time order (in a probabilistic fashion) regardless of whether realizations are dependent or independent from one another. This adds an important nuance to the common view that independent and identically distributed realizations imply the absence of time structure 
\parencites(see e.g.)()[][Section 1]{peters2013causal}[][Remark~6.7]{peters2017elements}.
In light of \cref{fig:sample_time}, the problems of nontemporal DAGs outlined in \cref{sec:ambiguity} and \cref{sec:acyclicity} can be understood as resulting from the lack of specifying the within-realization time points that variables correspond to.
The notion of time we aim to formalize in the remainder of this section is therefore within-realization.

\subsection{A Formal Notion of Time for Causal Variables}\label{sec:formalization}
Beginning with variables that refer to a quantity at a single time point, we develop our formalism to allow for composite variables, which can refer to quantities at multiple, possibly random, time points. 

\paragraph{Single time point}
In the simplest case, each variable refers to a quantity at exactly one time point relative to a realization's timeline.
By default, we take the time point $t$ of a variable $\mathbb{X}_t$ to be the time point at which it is measured, but other specifications are equally possible (e.g.\ proxy measurements of past quantities).
For simplicity, each within-realization timeline starts at $0$.
This formalization captures the notion of time employed in the literature on causality in time series.
In this simple model, the acyclicity assumption holds by the order of time.
To probe the acyclicity assumption where it may be violated and to model richer settings, we must allow for variables that can refer to quantities at multiple time points.

\paragraph{Multiple time points}
In all but the most strictly controlled settings, variables may refer to multiple and varying relative time points in different realizations. In observational medical data, for example, the time structure between measurements may vary from patient to patient depending on the availability of medical personnel and equipment.
To formalize such settings in the form of composite variables, we introduce additional objects. We index them with the same lower-case letter if they are used in the definition of the same composite variable.
Let $\Tcal_x \subset\mathbb{R}_+$ be a countable set of possible time points.
We assume that for each time point $t\in\Tcal_x$ there exists a corresponding real-valued random variable $\mathbb{X}_t$, which we capture collectively by the real-valued stochastic process $(\mathbb{X}_t)_{t\in\Tcal_x}$.
For a fixed integer $k_x\in\mathbb{N}^*$, let $\T_x\subseteq\Tcal_x$ be a random variable taking values in the set of $k_x$-element subsets of $\Tcal_x$.
The process $(\mathbb{X}_t)_{t\in\T_x}$ can then be used to formalize a random selection from multiple different time points.
In each realization of the stochastic process, the set of time points is drawn according to the law of $\T_x$. This law can be thought of as a law over measurement procedures.

\paragraph{Aggregating multiple time points}
Many variables of practical interest are aggregations of quantities at multiple time points, e.g.\ in the form of averages over time or differences between time points. For a prominent example, consider the historical debate regarding the effect of smoking on lung cancer \parencite[see e.g.][Section 9.5]{spirtes2001causation}. In this debate, \textit{smoking} refers to a composite variable that describes a behavioral pattern over a time period.
To formalize such settings, let $f_x\colon \mathbb{R}^{k_x}\to\mathbb{R}$
be a fixed real-valued aggregation function such that
$f_x(\mathbf{x})\neq f_x(\mathbf{x'})$ if $\mathbf{x},\mathbf{x'}\in\mathbb{R}^{k_x}$ differ in any single component. This ensures that an intervention on one temporal component also results in a change in the aggregate variable. 
In the smoking-cancer example, a real-valued smoking variable might, for instance, be the difference in total cigarettes smoked between two time points divided by the length of the time period.

\paragraph{Putting it all together}
On the basis of the objects introduced and summarized in \cref{tab:objects}, we can now formalize a composite variable $\Xt$ that can refer to a quantity at a single time point, or to a random selection or aggregation of quantities at multiple time points.
\begin{table}[H]
    \small
    \centering
    \renewcommand{\arraystretch}{1.2}
    \begin{tabular}{|l|l|l|}
        \hline
        \textbf{Function} & \textbf{Object} & \textbf{Interpretation}\\
        \hline
        Single time point & $\mathbb{X}_t$ & Single variable at time point $t$\\
        \hline
        & $k_x$ & Number of time points per realization of $\Xt$\\
        Multiple time points  & $\Tcal_x$ & Set of possible time points of which $\T_x$ is a random subset\\
        & $\T_x$ & Actual time points; random subset of $k_x$ time points of $\Tcal_x$\\
        \hline
        Aggregating time points & $f_x$ & Aggregation function of $(\mathbb{X}_t)_{t\in\T_x}$\\
        \hline
    \end{tabular}
    \caption{The objects used for formalizing a composite variable $\Xt$ with an explicit notion of time.}
    \label{tab:objects}
\end{table}
\noindent
For any valid $k_x$ and any distribution of $\T_x$, we define the composite causal variable
\begin{align}
    X_{\T_x}\coloneq f_x\left((\mathbb{X}_t)_{t\in\T_x}\right),
    \label{eq:temporal_causal_var}
\end{align}
which explicitly references corresponding time points in contrast to its nontemporal counterpart $X$.
Owing to the essential role of time in causality and the problems due to its absence in conventional causal DAGs outlined in \cref{sec:ambiguity,sec:acyclicity}, we choose to keep the time index explicit even after the aggregation into a single real-valued random variable.
The objects in \cref{tab:objects} encompass a wide range of possible variable definitions, though in principle our formalism allows for further generalization, for example by making $k_x$ a random variable. 
Moving forward, all causal variables will be composite variables of the form given in \cref{eq:temporal_causal_var}.

\paragraph{The atomic causal DAG}
Following the time series DAG literature \parencite[e.g.][]{hyttinen2012,bongers2021foundations}, we assume the existence of a causal DAG between variables that refer to quantities at single time points.
We refer to this DAG as the \textit{atomic} causal DAG, and within it all causes precede their effects (see \cref{graydef:cause_prec_effect}).
From the atomic causal DAG it is possible to derive a directed graph between composite variables, which may or may not be acyclic.
Interventions on $\Xt$ may target $\X_{t}$ at any time point $t$ that can appear with positive probability in $\T_x$.
We say that $\Xt$ causes another causal variable $\Yt$ if there exists $(t_x, t_y)$ that occurs jointly with positive probability and $\mathbb{X}_{t_x}$ causes $\mathbb{Y}_{t_y}$ in the atomic DAG, without considering self-loops \parencite[see \cref{graydef:causation}; for a detailed treatment of inference rules over cluster variables we refer to][]{anand2023causal}.
We denote the existence of a causal path between two variables by $\leadsto$. If the causal path consists of a single edge, then $\leadsto$ can be replaced by a straight arrow $\to$.

\paragraph{Running example}
To illustrate how time point-specific variables relate to composite causal variables, consider a generic medical treatment-symptom setting following the formalization introduced above with a treatment process $(\mathbb{X}_t)_{t\in\Tcal_x}$ with $\Tcal_x=\{0,6\}$, and a symptom process $(\mathbb{Y}_t)_{t\in\Tcal_y}$ with $\Tcal_y=\{4,10\}$.
The timeline is shown in \cref{fig:atomic}.
For visual clarity, the temporal components of each composite variable are drawn on separate lines belonging to the same within-realization timeline, as indicated by the vertical connection.
The dotted arrows show causal relationships that are compatible with the time order (shown only in this figure).
\begin{figure}[H]
    \centering
    \begin{subfigure}{\linewidth}
        \centering
        \begin{tikzpicture}[scale=1]
            \draw[blue!20,thick] (0,0) -- (10.5,0);
            \pgfmathsetmacro{\d}{-.6};
            \pgfmathsetmacro{\dtime}{-1};
            \draw[gray,thick] (0,0.2) -- (0*10,\dtime-0.1);
            \draw[red!20,thick] (0,\d) -- (10.5,\d);
            \draw[->,gray,thick] (0,\dtime) -- (10.5,\dtime);
            \draw[gray,thick] (4,\dtime+0.1)--(4,\dtime-0.1);
            \draw[gray,thick] (6,\dtime+0.1)--(6,\dtime-0.1);
            \draw[gray,thick] (10,\dtime+0.1)--(10,\dtime-0.1);
            \node at ( 0,\dtime-0.4) {$0$};
            \node at ( 4,\dtime-0.4) {$4$};
            \node at ( 6,\dtime-0.4) {$6$};
            \node at (10,\dtime-0.4) {$10$};
            \node[blue] (x1) at (0,0) {$\tick$};
            \node[blue] (x2) at (6,0) {$\tick$};
            \node at (0,0.4) {$\mathbb{X}_{0}$};
            \node at (6,0.4) {$\mathbb{X}_{6}$};
            \node[red] (y1) at ( 4,\d) {$\tick$};
            \node[red] (y2) at (10,\d) {$\tick$};
            \node at ( 4,\d+0.35) {$\mathbb{Y}_{4}$};
            \node at (10,\d+0.35) {$\mathbb{Y}_{10}$};
            \draw[->,dotted] (x1).. controls (1,.5) and (5,.5) ..(x2);
            \draw[->,dotted] (x1)--(y1);
            \draw[->,dotted] (x2)--(y2);
            \draw[->,dotted] (y1)--(x2);
            \draw[->,dotted] (y1).. controls (5,\d-.2) and (9,\d-.2) ..(y2);
            \draw[->,dotted] (x1).. controls (1,\d+1.4) and (7.5,\d+1.9) ..(y2);
            \node (t) at (11.2,-1) {time};
        \end{tikzpicture}
        \label{fig:ex1_atomic}
    \end{subfigure}
    \caption{Atomic causal DAG.}
    \label{fig:atomic}
\end{figure}
On the basis of the stochastic processes, one can define composite variables $\Xt$ and $\Yt$. 
For instance, one may define the composite treatment $\Xt\coloneq(\mathbb{X}_0+\mathbb{X}_6)/2$, and the composite outcome $\Yt$ as the identity evaluated either at $\mathbb{Y}_4$ or $\mathbb{Y}_{10}$ with equal probability. \cref{tab:generic_ex} below summarizes this in terms of the objects in \cref{tab:objects}. For more examples, see the forthcoming \cref{sec:examples}.
\begin{table}[H]
    \centering
    \begin{tabular}{c|c|c|c|c}
        & $\Tcal$ (possible) & $k$ & $\T$ (actual) & $f$\\
        \hline
        $\Xt$ & $\{0, 6\}$ & $2$ & $\{0, 6\}$ & average\\
        $\Yt$ & $\{4, 10\}$ & $1$ & $\{4\}$ or $\{10\}$ & identity
    \end{tabular}
    \caption{Composite variable definitions.}
    \label{tab:generic_ex}
\end{table}

\subsection{A Tale of Two Acyclicities}\label{sec:two_acyclicities}

The temporality of the variables in a causal DAG plays a central role in the justification of the acyclicity assumption.
We partition the acyclicity assumption into two separate and individually sufficient conditions with distinct characteristics: acyclicity by a time order that prevents cycles, and acyclicity by the absence of causal cycles where the time order allows them. 
The latter is a stronger assumption that requires additional justification.
A detailed formalization of the concepts introduced here can be found in \cref{app:formalization}.
We begin by defining \textit{time-acyclicity} as acyclicity by the time order. It is the reason for acyclicity in many common examples in the literature 
\parencites(see e.g.)()[][]{spirtes2001causation}{pearl2009causality}[][Chapter 14]{spohn2012laws}.
\begin{mydefinition*}[Time-acyclicity (see \cref{graydef:time_acyclicity} for a formal version)]
    Time-acyclicity holds between two causal variables if and only if the latest time point of one always precedes the earliest time point of the other.
\end{mydefinition*}
\noindent
Discussions of the acyclicity assumption typically focus on the presence or absence of causal cycles such as feedback loops \parencite[see e.g.][]{spirtes2000constructing,brouillard2024landscape}. 
Causal cycles can only occur if there is no time-acyclicity, which leads us to the definition of \textit{effect-acyclicity} as acyclicity by the absence of possible causal relationships that would lead to cycles.

\begin{mydefinition*}[Effect-acyclicity (see \cref{graydef:effect_acyclicity} for a formal version)]
    Effect-acyclicity holds between two causal variables if and only if time-acyclicity does not hold for them, yet there is still no causal cycle between them.
    \label{def:effect_acyclicity}
\end{mydefinition*}
\noindent
A special case of effect-acyclicity between two causal variables is given if there is no way for one to affect the other, regardless of the temporal relationship between them. We refer to this condition as \textit{total effect-acyclicity} and define it as follows.
\begin{mydefinition*}[Total effect-acyclicity (see \cref{graydef:total_effect_acyclicity} for a formal version)]
    Total effect-acyclicity holds between two causal variables if and only if there is no causal cycle between them regardless of which of their possible time points are selected.
    \label{def:effect_acyclicity}
\end{mydefinition*}
\noindent
Total effect-acyclicity is the type of acyclicity with the most far-reaching consequences. For a nonempty graph, total effect-acyclicity describes a generic asymmetry irrespective of the definition of a particular composite variable. Such asymmetries are a key reason for the appeal of causal models in other fields, in particular in machine learning \parencite{scholkopf2019causality}.
Unfortunately, they may be hard to justify since they are statements about the entire underlying stochastic processes (compare the formal definitions in \cref{app:formalization}).

\paragraph{The contrast to acyclicity for nontemporal variables}
In our formalism, the correspondence of causal variables to time points is explicit. Therefore, an acyclic causal relationship between two variables such as $\Xt\leadsto \Yt$ does not in general preclude that some $\mathbb{Y}_{t}$ may cause some $\mathbb{X}_{t'}$.
Note the contrast to the paradigm of nontemporal DAGs, in which a path $X\leadsto Y$ categorically excludes any path $Y\leadsto X$, but it remains unclear if this is meant to refer to time- or effect-acyclicity for a particular sets of time points, or to a fundamental asymmetry in the sense of total effect-acyclicity.

\section{Illustrative Examples}\label{sec:examples}

We begin by discussing a series of examples illustrating our formalism and how to justify acyclicity using the concepts presented in \cref{sec:two_acyclicities}.
We use our running example from \cref{fig:atomic} consisting of treatment process $(\mathbb{X}_t)_{t\in\Tcal}$ with $\Tcal_x=\{0,6\}$ and symptom process $(\mathbb{Y}_t)_{t\in\Tcal_y}$ with $\Tcal_y=\{4,10\}$.
Then, we show how our formalism can be used to reason systematically about causal cycles, time scales, and unobserved confounding.
Finally, we demonstrate how it may help uncover faithfulness violations.

\begin{example}{ (Single time point)}
\label{ex:selection}
For our first and simplest scenario, we define our composite variables to be a quantity at a single one of all possible time points.
For instance, let $\Xt\coloneq\X_{0}$ and $\Yt\coloneq\Y_{10}$, with $\T_x=\{0\}$, $\T_y=\{10\}$, $k_x=k_y=1$, and both of $f_x$ and $f_y$ defined as the identity function of the variable in their singleton input set.
This is an example of the time point-specific variable definition also used in the literature on causality in time series.
The example exhibits time-acyclicity, as is easy to see since the time order allows $\Xt$ to be a cause of $\Yt$, but not vice versa.
\end{example}

\begin{example}{ (Multiple time points)}
\label{ex:mixing}
A more complex scenario emerges for variables that can refer to a quantity at a randomly chosen time point.
For instance, let $\Xt$ be equal to either $\X_{0}$ or $\X_{6}$, and $\Yt$ be equal to either $\Y_{4}$ or $\Y_{10}$.
This is the case if $\T_x$ is either $\{0\}$ or $\{6\}$ with positive probability, and $\T_y$ is either $\{4\}$ or $\{10\}$ with positive probability, $k_x=k_y=1$, and $f_x$ and $f_y$ are the identity as before.
Here, time-acyclicity is no longer as easy to justify since, by the time order, $\Xt$ could cause $\Yt$ and vice versa, e.g.\ if $\mathbb{X}_0\leadsto\mathbb{Y}_4$ and $\mathbb{Y}_4\leadsto\mathbb{X}_6$.
Still, time-acyclicity may hold, e.g.\ if the joint event of $(\T_y=\{4\},\T_x=\{6\})$ has probability $0$, meaning that the symptom measurement always follows the treatment.
If all combinations of time points have positive probability, acyclicity requires justification via effect-acyclicity, e.g.\ by the absence of a causal effect of $\mathbb{Y}_4$ on $\mathbb{X}_6$ in the atomic DAG.

\end{example}

\begin{example}{ (Aggregation)}
\label{ex:aggregation}
Variables of interest may be aggregates of quantities at multiple time points.
For instance, let $\Xt=(\X_0+\X_6)/2$ and $\Yt=(\Y_4+\Y_{10})/2$, with $\T_x=\{0,6\}$ and $\T_y=\{4,10\}$, $k_x=k_2=2$, and $f_x$ and $f_y$ taking the average of their inputs.
In this setting, the definition of $\T_x$ and $\T_y$ rules out time-acyclicity.
The acyclicity assumption can only hold by effect-acyclicity, otherwise the variables cannot be validly represented in a causal DAG.
Provided there is no causal cycle, the time order allows for any causal DAG between $\Xt$ and $\Yt$. A causal DAG including the path $\Xt\leadsto\Yt$ would require that $\mathbb{Y}_{4}$ does not cause $\mathbb{X}_{6}$, as in a non-adaptive medical treatment regime.
Analogously, a causal DAG including the path $\Yt\leadsto\Xt$ would require that $\mathbb{X}_{0}$ does not cause either of $\mathbb{Y}_{4}$ and $\mathbb{Y}_{10}$, and that $\mathbb{X}_{6}$ does not cause $\mathbb{Y}_{10}$.
\end{example}

\paragraph{Total effect-acyclicity in the examples}
\cref{ex:selection,ex:mixing,ex:aggregation} also show why total effect-acyclicity is a strong assumption.
In making a statement about all possible composite variables, total effect-acyclicity establishes acyclicity for all of the examples simultaneously.
In settings with many interleaved time points this would likely be difficult, unless there is a fundamental asymmetry between the processes.
Such an asymmetry exists, for example, between \textit{solar flares} and \textit{power outages}. The former can cause the latter, but never vice versa, making total effect-acyclicity plausible.
However, this example also showcases that certainty about the absence of causal paths in one direction often comes at the cost of variables without physically well-defined interventions, in this case on the solar flares
\parencite[for more cause-effect pairs, see][]{mooij2016distinguishing}.
Finally, note that even given total effect-acyclicity, it is still necessary to specify the time of causal variables, since the edges still depend on the variables' temporal relationships.

\subsection{Defining Variables: Cycles, Time Scales, and Unobserved Confounding}\label{sec:utility}
The issues of causal cycles, appropriate time scales, and unobserved confounding are of general interest to causal modelling, but they are particularly pertinent when the temporal nature of the variables is taken into account.
Here, we illustrate how our formalism may be used to reason about them systematically.

Consider the interaction between the treatment \textit{caffeine consumption} and the symptom \textit{sleep} \parencite[cf.][]{stucky2024community}.
Both variables are typically defined as aggregates over measurements at several time points, and hence cannot neatly be time-stamped with a single time point.
Assume that, over a short period of time, sleep affects caffeine consumption and not vice versa, yet over a longer period, caffeine consumption also affects sleep.
For illustration, let $(\mathbb{X}_t)_{t\in\Tcal_x}$ be a stochastic process of sleep measurements (e.g.\ via brain activity) at time points $\Tcal_x=\{1,2\}$, and let $(\mathbb{Y}_t)_{t\in\Tcal_y}$ be a stochastic process of caffeine measurements at time points $\Tcal_y=\{0,1,10\}$.
Let the atomic causal DAG be that shown in \cref{fig:time_scale}.
\begin{figure}[H]
    \centering
    \begin{tikzpicture}[]
        \draw[blue!20,thick] (0,0) -- (10.5,0);
        \pgfmathsetmacro{\d}{-.6};
        \pgfmathsetmacro{\dtime}{-1};
        \draw[gray,thick] (0,0.2) -- (0*10,\dtime-0.1);
        \draw[red!20,thick] (0,\d) -- (10.5,\d);
        \draw[->,gray,thick] (0,\dtime) -- (10.5,\dtime);
        \draw[gray,thick] (1,\dtime+0.1)--(1,\dtime-0.1);
        \draw[gray,thick] (2,\dtime+0.1)--(2,\dtime-0.1);
        \draw[gray,thick] (10,\dtime+0.1)--(10,\dtime-0.1);
        \node at ( 0,\dtime-0.4) {$0$};
        \node at ( 1,\dtime-0.4) {$1$};
        \node at ( 2,\dtime-0.4) {$2$};
        \node at (10,\dtime-0.4) {$10$};
        \node[blue] (y1) at (1,0) {$\tick$};
        \node at ( 1,0.4) {$\X_{1}$};
        \node[blue] (y2) at (2,0) {$\tick$};
        \node at ( 2,0.4) {$\X_{2}$};
        \node[red] (x1) at (0,\d) {$\tick$};
        \node at (0,\d+0.3) {$\Y_{0}$};
        \node[red] (x2) at (1,\d) {$\tick$};
        \node at (1,\d+0.3) {$\Y_{1}$};
        \node[red] (x3) at (10,\d) {$\tick$};
        \node at (10,\d+0.3) {$\Y_{10}$};
        \draw[->] (x1)--(y1);
        \draw[->] (x2)--(y2);
        \draw[->] (x1)to[out=20,in=160](x2);
        \draw[->] (x2).. controls (2.5,\d+.2) ..(x3);
        \draw[->] (y1)to[out=20,in=160](y2);
        \draw[->] (y1).. controls (3,-.3) and (5,-.3) ..(x3);
        \draw[->] (y2)--(x3);
        \node (t) at (11.2,-1) {time};
  rectangle (current bounding box.north east);
    \end{tikzpicture}
    \caption{Atomic causal DAG.}
    \label{fig:time_scale}
\end{figure}

\paragraph{Unrolling causal cycles}
It is often suggested that cyclic causal relationships are acyclic at a sufficiently fine-grained temporal resolution 
\parencites(e.g.)()[][Section 1.4.1]{pearl2009causality}[][Section 2.3.3]{peters2017elements}, 
meaning one could unroll cycles in time to obtain a DAG.
Such an operation relies on a correspondence between variables and time points,
which is what our formalism makes explicit.\\
In the example, let for instance caffeine consumption be $\Xt\coloneq\X_2-\X_1$, and sleep be $\Yt\coloneq(\Y_0+\Y_1+\Y_{10})/3$.
This gives rise to a cycle on the level of composite variables, as shown in \cref{fig:composite_cycle}.
The causal cycle can be unrolled by splitting $\Yt$ into the pre-treatment variable $Y_{\T_{y_\text{pre}}}\coloneq(\Y_0+\Y_1)/2$ and the post-treatment variable $\mathbb{Y}_{\T_{y_\text{post}}}\coloneq\Y_{10}$.
In combination with $\Xt$, these new variables form the DAG in \cref{fig:resolution_graph}, which shows why the past of an outcome variable is a natural candidate for a confounder \parencite[cf.][Section 1.8]{imbens2015causal}.
\begin{figure}[H]
    \centering
    \begin{subfigure}{0.3\linewidth}
        \centering
        \begin{tikzpicture}
            \node (A) at (0,0) {$X_{\T_x}$};
            \node (B) at (2,0) {$Y_{\T_y}$};
            \node (C) at (0,.5) {$\phantom{Z}$};
            \draw[->] (A) .. controls +(0.5,0.5) and +(-0.5,0.5) .. (B);
            \draw[->] (B) .. controls +(-0.5,-0.5) and +(0.5,-0.5) .. (A);
            \node (phant) at (1,-0.55) {$\phantom{a}$};
        \end{tikzpicture}
        \caption{Cyclic causal graph.}
        \label{fig:composite_cycle}
    \end{subfigure}
    \hskip2em
    \begin{subfigure}{.3\linewidth}
        \centering
        \begin{tikzpicture}
            \node (C) at (1.5,.7) {$Y_{\T_{y_\text{pre}}}$};
            \node (A) at (0,0) {$X_{\T_{x}}$};
            \node (B) at (3,0) {$Y_{\T_{y_\text{post}}}$};
            \draw[->] (A) -- (B);
            \draw[->] (C) -- (A);
            \draw[->] (C) -- (B);
        \end{tikzpicture}
        \caption{Unrolled causal DAG.}
        \label{fig:resolution_graph}
    \end{subfigure}
    \caption{Resolving a causal cycle by unrolling in time.}
    \label{fig:unrolling}
\end{figure}
\noindent
Note that unrolling a causal cycle requires measurements at a higher temporal resolution and may change the interpretation of the model 
\parencites(cf.)()[][Section 1]{hyttinen2012}[][Section 1]{mooij2011causal}.
Composite variables in settings characterized by high-frequency interactions, such as stock prices or gene expressions, may therefore be difficult to unroll into meaningful DAGs.

\paragraph{Time scales and unobserved confounding}
Episodic settings such as those suggested by our medical examples provide a natural time scale, yet in general the time scale of causal models is not fixed and can affect their structure and properties \parencite[][]{weinberger2020representation}.
For instance, in the example on caffeine consumption and sleep shown in \cref{fig:time_scale,fig:unrolling}, one could focus on a short time scale by reducing the model to $\Y_{t_\text{pre}}$ and $\Xt$,
or focus a long time scale by reducing the model to $\Xt$ and $\Y_{t_\text{post}}$.
Though both of the resulting models are acyclic, the different time scales lead to opposite causal effects between the respective caffeine consumption and sleep variables. 
Moreover, in the long-term model there is confounding by $\Y_0$ and $\Y_1$, which highlights another important aspect of choosing an appropriate time scale.
Quantities at time points that are not sampled may act as unobserved confounders, and the quantities modelled may exist even before the time horizon of the causal model, which greatly complicates justifying the absence of unobserved confounding.
Though this has been noted in the context of time series \parencite{thams2024identifying}, our temporal perspective on causal variables highlights that quantities at past time points may act as unobserved confounders in any causal setting.

\subsection{Uncovering Faithfulness Violations}\label{sec:faithfulness}

The foundational idea for discovering causal DAGs from data is to test for conditional independencies between variables to find a Bayesian network which may be interpreted as the causal DAG or equivalence class containing it \parencite[][]{spirtes2001causation}.
This approach relies upon the causal faithfulness assumption 
\parencite[see e.g.][Definition~6.33]{peters2017elements}.
Using our formalism, we give an example to demonstrate that variables that refer to quantities at multiple time points may give rise to faithfulness violations that violate the time order. 
Faithfulness violations arising from variable groupings are described in \textcite[][Section 3]{parviainen2017} and further analyzed in \textcite[][Sections 5 and 6]{wahl2024foundations}.
This type of faithfulness violation cannot be easily dismissed by the arguments in \textcite[][Section 2.4]{pearl2009causality}, \textcite[][Section 3.5.2 and Theorem 3.2]{spirtes2001causation}, and \cite[][]{Meek1995faithfulness}.
Consider a variation of our earlier medical setting of treatment process $(\mathbb{X}_t)_{t\in\Tcal_x}$ with $\Tcal_x=\{0\}$, and symptom process $(\mathbb{Y}_t)_{t\in\Tcal_y}$ with $\Tcal_y=\{2, 8\}$.
For this example, take the treatment to be aspirin and the symptom to be headache.
Let further $(\Z_t)_{t\in\Tcal_z}$ with $\Tcal_z=\{10\}$ be a third process that measures absence from work.
Let $\Xt\coloneq\X_0$, $\Yt\coloneq(\Y_2+\Y_8)/2$, and $\Zt\coloneq\Z_{10}$.
Assume further that aspirin at time $0$ cures headache at time $2$, and that headache at time $8$ causes absence from work at time $10$, as illustrated in the atomic causal DAG in \cref{fig:atomic_v}. 
The temporal aggregation into composite variables leads to a pairwise causal dependence between $\Xt$ and $\Yt$ and between $\Yt$ and $\Zt$, but not between $\Xt$ and $\Zt$, meaning there is no transitivity.
Applying the orientation rules in \cite{meek1995orientation} yields the Bayesian network shown in \cref{fig:v_struc}, which would be found using for example the PC algorithm for causal discovery \parencite[][Section 5.4.2]{spirtes2001causation}.
The edge $\Zt\dasharrow\Yt$ in the Bayesian network violates the time order.
\begin{figure}[H]
    \centering
    \begin{subfigure}{0.55\linewidth}
        \centering
        \begin{tikzpicture}[scale=0.6]
            \pgfmathsetmacro{\dtwo}{-0.5};
            \pgfmathsetmacro{\dthree}{-1};
            \pgfmathsetmacro{\dtime}{-1.5};
            \pgfmathsetmacro{\Tone}{2};
            \pgfmathsetmacro{\Ttwo}{8};
            \pgfmathsetmacro{\labeld}{0.4};
            \draw[->,thick,lightgray](0,\dtime)--(10.5,\dtime);
            \draw[thick,lightgray] (0,0.2) -- (0,\dtime-0.1);
            \draw[thick,lightgray] (2,\dtime+0.1) -- (2,\dtime-0.1);
            \draw[thick,lightgray] (8,\dtime+0.1) -- (8,\dtime-0.1);
            \draw[thick,lightgray] (10,\dtime+0.1) --(10,\dtime-0.1);
            \draw[blue!20,thick] (0,0) -- (10.5,0);
            \draw[green!20,thick] (0,\dtwo) -- (10.5,\dtwo);
            \draw[red!20,thick] (0,\dthree) -- (10.5,\dthree);
            \node (zero) at (0,\dtime-0.5) {$0$};
            \node (two) at (\Tone,\dtime-0.5) {$2$};
            \node (eight) at (\Ttwo,\dtime-0.5) {$8$};
            \node (ten) at (10,\dtime-0.5) {$10$};
            \node[blue] (x1) at ( 0,0) {$\tick$};
            \node at (0,0.45) {$\mathbb{X}_{0}$};
            \node[green!50!black] (z) at (10,\dtwo) {$\tick$};
            \node at (10,\dtwo+0.4) {$\mathbb{Z}_{10}$};
            \node[red] (y1) at (\Tone,\dthree) {$\tick$};
            \node[red] (y2) at (\Ttwo,\dthree) {$\tick$};
            \node at (2,\dthree+0.4) {$\mathbb{Y}_{0.2}$};
            \node at (8,\dthree+0.4) {$\mathbb{Y}_{0.8}$};
            \draw[->] (x1)--(y1);
            \draw[->] (y2)--(z);
        \end{tikzpicture}
        \caption{Atomic DAG.}
        \label{fig:atomic_v}
    \end{subfigure}
    \hfill
    \begin{subfigure}{0.44\linewidth}
        \centering
        \begin{tikzpicture}
            \node (x1) at (0,0) {$X_{\T_{x}}$};
            \node (x2) at (3,0) {$Z_{\T_{z}}$};
            \node (y)  at (1.5,-1) {$Y_{\T_{y}}$};
            \draw[->,dashed] (x1)--(y);
            \draw[->,dashed] (x2)--(y);
        \end{tikzpicture}
        \caption{Bayesian network of composite variables.}
        \label{fig:v_struc}
    \end{subfigure}
    \caption{Faithfulness violations can lead to Bayesian network edges pointing backward in time.}
    \label{fig:time_v}
    \label{fig:v_structure}
\end{figure}
This can be explained by a faithfulness violation. By \cref{graydef:causation}, the composite variables form the DAG \mbox{$\Xt\to\Yt\to\Zt$}.
The variables $\Xt$ and $\Zt$ are independent but not $d$-separated in that DAG.
The example highlights that interpreting Bayesian network edges causally means reaching beyond their formal semantics, which requires careful justification \parencite[Section 4.3]{dawid2010beware}.
\section{Implications for Statistical Causality}

The formalism presented in this paper aims to build upon the standard causal DAG framework. Nonetheless, our explicitly temporal composite variables mark a conceptual departure with far-reaching implications.

\paragraph{Causal discovery and causal representation learning}\label{sec:discussion}
The field of causal discovery aims to infer causal graphs, usually DAGs or their equivalence classes, from data using statistical methods \parencite[see e.g.][]{kitson2023survey}.
If one adopts our temporal perspective, this task suffers from an existential problem: without prior knowledge about the causal order or the atomic DAG, how is one to know if an acyclic causal graph exists between the variables in the first place?
For this reason, we consider the incorporation of time into causal DAGs a crucial step in identifying settings in which causal discovery can hope to succeed, and thus to address the lack of realism in causal discovery benchmarking \parencite[see e.g.][]{gentzel2019case,reisach2021beware,reisach2023scale,brouillard2024landscape}.
This implies a shift in perspective, away from discovering a causal order and toward treating it as necessary background knowledge. \parencite[cf.][]{stern2021causal, bang2023we}.
In the context of learning causal representations and their corresponding DAGs from data \parencite{scholkopf2021toward}, our contribution highlights the importance of taking the time dimension of causal representations into account \parencite[cf.][]{weinberger2020representation}.

\paragraph{Causal effect estimation and causally robust prediction}
A central benefit of causal DAGs is the do-calculus \parencite{pearl1995causal}, which enables assessing identifiability and selecting suitable control variables for a causal effect of interest.
However, the temporal relationship between variables may already provide useful information about suitable control variables, which explains why the potential outcomes framework enjoys great popularity for causal inference in the
applied sciences \parencite[see e.g.][]{angrist2009mostly,imbens2015causal}, seemingly without a need for causal DAGs.
Incorporating time into causal DAGs is likely to help outline the advantage they can offer more clearly.
Taking time into account also provides a new perspective on the task of causally robust prediction \parencite{buehlmann2020invariance}, and the promise of causal DAGs for machine learning \parencite[][]{scholkopf2019causality}.
In so far as variables correspond to single time points, causally robust prediction essentially amounts to predicting in the time order using quantities at the directly preceding causally relevant time points.
However, many variables of interest may be composite variables with cyclic causal relationships, which could explain why the practical usefulness of causal machine learning remains to be demonstrated \parencite{nastl2024causalpredictors}.

\paragraph{The applicability of DAGs as causal models}
Finally and perhaps most importantly, taking into account the role of time means tackling the question of the applicability of causal DAGs to real-world data in earnest.
Before the question \textit{what is the right DAG?} comes the question \textit{is there a DAG?}
Treating causal variables as instantiations highlights the ``statistical fantasy'' \parencite[][Chapter 6]{spirtes1991algorithm} of assuming that all confounders have been observed, since this assumption may require accounting for past instantiations in an infinite regress.
Moreover, the explicit role of time in our formalism clarifies that the validity of the acyclicity assumption is not self-evident.
A justification via time-acyclicity requires an analysis of the measurement procedure, for example by using background knowledge or statistical inference to characterize the law governing measurement time points. 
A justification via effect-acyclicity requires an analysis of the causal relationships in the atomic DAG.
Many variables of interest may not be amenable to causal analysis using DAGs, and what causal DAGs there are, are of a time-specific nature allowing for less sweeping conclusions.
In this light, the crucial step in applying causal DAGs consists in defining meaningful variables that allow for their use, rather than treating a set of variables as given and reasoning about \textit{the} DAG between them.
Incorporating time into causal DAGs therefore offers the potential for a more pragmatic approach to causal inference in line with the practices of applied research.
\section{Conclusion}\label{sec:conclusion}
Real-world quantities and their measurements exist and change in time.
If causes are to precede effects, a realization of a causal model consists in a time series.
It follows that causal relationships are relative with respect to the within-realization time structure.
For this reason, we argue that it is time (pun intended) to break with the tradition of nontemporal causal DAGs and incorporate time into the definition of variables in causal DAGs by default.
This perspective reveals that causal DAGs do not relate generic concepts but temporal instantiations, which is essential for their interpretation and reconciles an important difference between causal analysis based on DAGs versus causal analysis based on potential outcomes.
Variables that refer to quantities at multiple time points may give rise to causal cycles.
In order to establish the validity of the acyclicity assumption, one must show that the variables refer to quantities at time points following an order that prevents causal cycles (time-acyclicity), or that there are no causal cycles if the time order allows them (effect-acyclicity).
Incorporating time into causal DAGs is therefore essential for evaluating their applicability and interpretation as causal models.
Our formal treatment of time allows for doing so in a principled fashion and opens many questions for future research.
For example, it will be interesting to examine the change of causal effects with time distance, and to what extent some time points may be representative of others.
Moreover, further research is required to analyze the implications for arguments regarding the prevalence of faithfulness violations, and to apply our notion of time to other classes of graphical causal models.
Although our contribution shows the scope of causal DAGs to be more specific than it may otherwise appear, it also facilitates their discovery, clarifies their interpretation, and leaves them a less metaphysical and all the more practical tool for causal analysis.
\ifanonymous
\else
    \subsection*{Acknowledgements}
We thank Niels Richard Hansen (University of Copenhagen) for helping to inspire this project, as well as Emilie Devijver (Université Grenoble Alpes), Pouya Babakhani (University of Bath), and especially Isabelle Drouet (Sorbonne Université) for helpful discussions and feedback throughout the stages of writing.
AGR received funding from the European Union's Horizon 2020 research and innovation programme under the Marie Skłodowska-Curie grant agreement No 945332 and travel support under No 951847 \raisebox{-.05em}{\includegraphics[height=.85em]{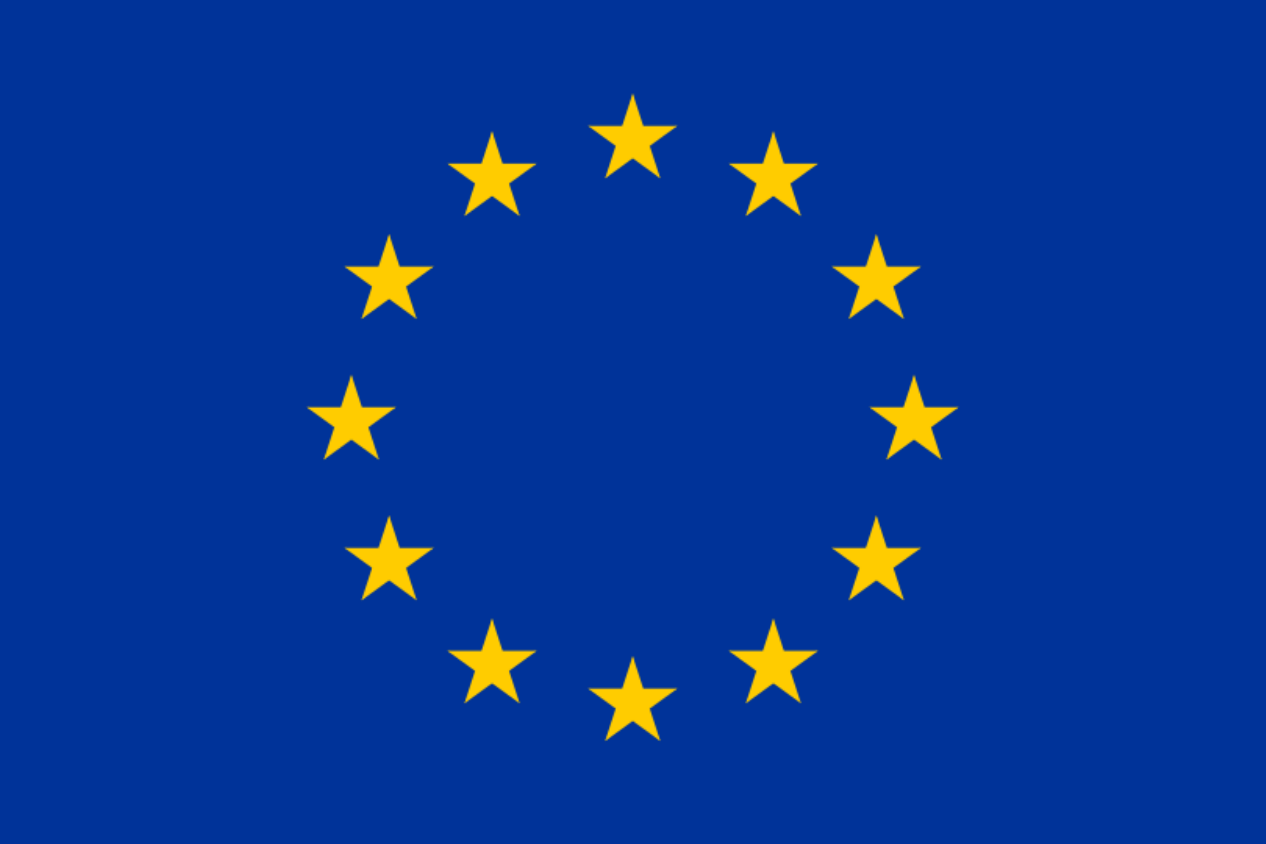}},
as well as mobility funding from Université Paris Cité via the 2024 BDMI program.
\fi

\printbibliography

\appendix
\crefalias{section}{appendix}
\newpage
\section{Formal Definitions}\label{app:formalization}

In this section we provide additional details regarding our formalism and definitions introduced in \cref{sec:formal_time}.
For convenience, we will use the following notation: for any $k\in\mathbb{N}^*$, for any set $S$ of cardinality greater than or equal to $k$, $\P_k(S)$ denotes the collection of subsets of $S$ which are of cardinality $k$.
We denote the support of the distribution of $\T_x$ as $\S_x\subseteq\P_{k_x}(\Tcal_x)$. 
We denote the support of the joint distribution of two sets of time points $\T_x$ and $\T_y$ with corresponding $\Tcal_x$ and $\Tcal_y$ as 
\begin{align}
    \S_{xy}\subseteq \S_x\times\S_y\subseteq\P_{k_x}(\Tcal_x)\times\P_{k_y}(\Tcal_y).
    \label{eq:joint_support}
\end{align}
Furthermore, we denote the temporal precedence of $\Xt$ with respect to $\Yt$ by the shorthand
\begin{align}
    &\T_x\prec \T_y \coloneq \forall(s_x, s_y)\in\S_{xy}, \max(s_x) < \min(s_y).
    \label{eq:time_order}
\end{align}

\begin{graydefinition}{Causes precede effects}{cause_prec_effect}
    We say that causes precede effects 
    if and only if, in the atomic causal DAG between the temporal components of any pair of stochastic processes $(\mathbb{X}_t)_{t\in\Tcal_x}$ and $(\mathbb{Y}_t)_{t\in\Tcal_y}$,
    \begin{align*}
        \forall t_x\in\Tcal_x,\forall t_y\in\Tcal_y,\left(\mathbb{X}_{t_x}\leadsto\mathbb{Y}_{t_y} \implies t_x < t_y\right).
    \end{align*}
\end{graydefinition}

\begin{graydefinition}{Causation for composite variables}{causation}
    A variable $\Xt$ causes another causal variable $\Yt$ if and only if there exists $(s_x,s_y)\in\S_{xy}$ with $t_x\in s_x$ and $t_y\in s_y$ such that $\mathbb{X}_{t_x}$ causes $\mathbb{Y}_{t_y}$ in the atomic DAG (we do not consider self-loops on the level of composite variables, e.g.\ if $\mathbb{X}_{t_1}$ causes $\mathbb{X}_{t_2}$ for some $t_1,t_2\in s_x$ for $s_x\in\S_x$).
\end{graydefinition}

\begin{graydefinition}{Acyclicity}{acyclicity}
    Acyclicity between two causal variables $X_{\T_x}$ and $Y_{\T_y}$ holds if and only if in the atomic causal DAG
    \begin{align*}
        &\exists(s_x,s_y)\in\S_{xy},\exists t_x\in s_x,\exists t_y\in s_y,(\mathbb{X}_{t_x}\leadsto\mathbb{Y}_{t_y})\implies\\
        &\forall(s'_x,s'_y)\in\S_{xy},\forall t'_x\in s'_x,\forall t'_y\in s'_y,\lnot(\mathbb{Y}_{t'_y}\leadsto\mathbb{X}_{t'_x}).
    \end{align*}
    We further call a graph $\G$ acyclic if acyclicity holds for all pairs of variables in $\G$.
\end{graydefinition}

\begin{graydefinition}{Time-acyclicity}{time_acyclicity}
    Time-acyclicity between two causal variables $X_{\T_x}$ and $Y_{\T_y}$ holds if and only if
    \begin{align*}
        (\T_x \prec \T_y) \lor (\T_y\prec \T_x).
    \end{align*}
    We further call a graph $\G$ time-acyclic if time-acyclicity holds for all pairs of variables in $\G$.
\end{graydefinition}

\begin{graydefinition}{Effect-acyclicity}{effect_acyclicity}
    Effect-acyclicity holds between two causal variables $X_{\T_x}$ and $Y_{\T_y}$ if and only if there is no time-acyclicity by \cref{graydef:time_acyclicity}, that is,
    \begin{align*}
        \lnot\left((\T_x\prec \T_y) \lor (\T_y\prec \T_x)\right),
    \end{align*}
    yet there is acyclicity by \cref{graydef:acyclicity}.
    We further call a graph $\G$ effect-acyclic if effect-acyclicity holds for all pairs of variables in $\G$.
\end{graydefinition}

\begin{graydefinition}{Total effect-acyclicity}{total_effect_acyclicity}
    Total effect-acyclicity between two real-valued stochastic processes $(\mathbb{X}_t)_{t\in\Tcal_x}$ and $(\mathbb{Y}_t)_{t\in\Tcal_y}$
    holds if and only if in the atomic causal DAG
    \begin{align*}
        &\exists(t_x,t_y)\in\Tcal_x\times\Tcal_y\text,(\mathbb{X}_{t_x}\leadsto\mathbb{Y}_{t_y})
        \implies \forall(t'_x,t'_y)\in\Tcal_x\times\Tcal_y,\lnot(\mathbb{Y}_{t'_y}\leadsto\mathbb{X}_{t'_x}).
    \end{align*}
    In this case, we say that total effect-acyclicity holds for any $\Xt$ and $\Yt$ defined with respect to $(\mathbb{X}_t)_{t\in\Tcal_x}$ and $(\mathbb{Y}_t)_{t\in\Tcal_y}$, respectively.
    We further call a graph $\G$ totally effect-acyclic if effect-acyclicity holds for all pairs of variables in $\G$.
\end{graydefinition}

\end{document}